\documentclass[aps, prd, amsmath, floats, floatfix, twocolumn, nofootinbib, superscriptaddress, showpacs]{revtex4-1}
\usepackage{graphicx}
\usepackage{color}
\usepackage{latexsym}

\newcommand{\be}{\begin{eqnarray}}
\newcommand{\ee}{\end{eqnarray}}

\begin{document}

\title{Testing the Kerr nature of black hole candidates using iron line reverberation mapping in the CPR framework}

\author{Jiachen Jiang}
\affiliation{Center for Field Theory and Particle Physics and Department of Physics, Fudan University, 200433 Shanghai, China}

\author{Cosimo Bambi}
\email[Corresponding author: ]{bambi@fudan.edu.cn}
\affiliation{Center for Field Theory and Particle Physics and Department of Physics, Fudan University, 200433 Shanghai, China}
\affiliation{Theoretical Astrophysics, Eberhard-Karls Universit\"at T\"ubingen, 72076 T\"ubingen, Germany}

\author{James F. Steiner}
\altaffiliation{Einstein Fellow}
\affiliation{MIT Kavli Institute, Cambridge, MA 02139, United States}

\date{\today}

\begin{abstract}
The iron K$\alpha$ line commonly observed in the X-ray spectrum of black hole candidates is produced by X-ray fluorescence of the inner accretion disk. This line can potentially be quite a powerful tool to probe the spacetime geometry around these objects and test the Kerr black hole hypothesis. In a previous paper, we studied the ability to constrain possible deviations from the Kerr solution from the standard time-integrated iron line spectrum within the Cardoso-Pani-Rico framework. In the present work, we expand on that study and consider iron line reverberation mapping in the CPR framework.  That is, we consider the time-evolution of the iron line profile in response to fluctuations in the X-ray primary source. Our simulations clearly show that the time information in reverberation mapping can better constrain the background metric than the time-integrated approach, and this is true, notably, for the deformation parameter $\epsilon^r_3$, which is only weakly informed by a time-integrated observation.
\end{abstract}


\maketitle


\section{Introduction \label{s-1}}

In the past 100 years, general relativity has successfully passed a large number of tests and at present there are no clear indications of deviations from its predictions~\cite{will}. However, the theory has been mainly tested in weak gravitational fields, in particular with experiments in the Solar System and observations of binary pulsars. Signatures of new physics can more likely manifest in more extreme conditions. The ideal laboratory to test general relativity in strong gravitational fields is the spacetime around astrophysical black hole (BH) candidates.

BHs are expected to be the final product of gravitational collapse~\cite{collasso} (e.g., for the most massive stars) and we  thus expect a large population of BHs in the Universe. In 4-dimensional general relativity, the only uncharged BH solution is described by the Kerr metric, which has only two parameters, namely the mass $M$ and the spin angular momentum $J$ of the BH. A fundamental limit for a Kerr BH is the bound $|a_*| \le 1$, where $a_* = a/M = J/M^2$ is the BH spin parameter. For $|a_*| > 1$ the Kerr metric describes a naked singularity and there are arguments to believe that such objects cannot exist~\cite{enrico}\footnote{We note that such a bound does not hold in other spacetimes. In particular, it is possible to create non-Kerr compact objects with $|a_*| > 1$~\cite{ss} and overspin some non-Kerr BHs, which destroys their horizons~\cite{eh}.}. Hairy BHs are possible in 4-dimensional general relativity, but this requires the presence of exotic fields~\cite{v2-hr14}.

Observationally, there is robust evidence for the existence of two classes of BH candidates~\cite{narayan}: stellar-mass BH candidates in X-ray binaries and supermassive BH candidates in galactic nuclei. These objects are termed ``candidates'' because while they are naturally interpreted as BHs, there is no direct observational confirmation of their BH nature. More precisely, stellar-mass BH candidates are too compact and massive to be neutron stars~\cite{ruffini}, while supermassive BH candidates are too massive, compact, and old to be clusters of neutron stars~\cite{maoz}. In the framework of standard physics, both object classes should be the Kerr BHs of general relativity, and they could be something else only in the presence of new physics. Observational confirmation of the Kerr BH hypothesis can thus be seen as a fundamental test of general relativity in the regime of strong gravity.

Electromagnetic radiation emitted by the gas in the inner part of the accretion disk can be exploited to test the nature of BH candidates and constrain possible deviations from the Kerr metric~\cite{review}. However, this is observationally quite challenging, in part owing to parameter degeneracy. The thermal spectrum of thin accretion disks~\cite{cfm1,cfm2} has a simple shape, and, in general, it is difficult to measure the spin parameter while simultaneously constraining possible deviations from the Kerr solution~\cite{cfm3}. The iron K$\alpha$ technique~\cite{iron1,iron2} is potentially more powerful, but the limitation preventing current observations from being used to test the Kerr paradigm is a combination of the limited signal strength and the degree of sophistication of present models~\cite{jjc1,jjc2}. Current data can exclude that BH candidates are exotic dark stars~\cite{exotic1} or traversable wormholes~\cite{exotic2}, because the iron line does not match in the available data. However, many less exotic scenarios are quite challenging to rule out.

The aim of this paper is to extend previous work in the literature and investigate the advantages, if any, of iron line reverberation with respect to the standard time-integrated line spectrum for testing the Kerr metric in the Cardoso-Pani-Rico (CPR) parametrization~\cite{cpr}. In a reverberation measurement, the iron line signal is detected as a function of time in response to an instantaneous emission of radiation from the corona, which acts as the primary source of hard X-rays. The resulting line spectrum as a function of both time and across photon energy is called the 2D transfer function. The point is that the time information in the measurement enables one to distinguish radiation generated from different parts of the accretion disk, and this helps to better probe the relativistic effects occurring in the vicinity of the compact object.

The constraining capability of the time-integrated (spectral) approach on the CPR deformation parameters $\epsilon^t_3$ and $\epsilon^r_3$ was studied in Ref.~\cite{jjc2}. In the non-rotating limit, $\epsilon^t_3$ only alters the metric coefficient $g_{tt}$, while $\epsilon^r_3$ only the metric coefficient $g_{rr}$. In the case of a rotating BH, it is still true that $g_{tt}$ is only altered by $\epsilon^t_3$ and that $g_{rr}$ is only altered by $\epsilon^r_3$, but now both the deformation parameters enter $g_{t\phi}$ and $g_{\phi\phi}$ in a non-trivial way. We  found that the CPR deformation parameter $\epsilon^t_3$ is relatively easy to constrain with a high-count spectrum, while $\epsilon^r_3$ is much more elusive and even the iron line of a fast-rotating Kerr BH may be mimicked by a slow-rotating non-Kerr object with a large positive $\epsilon^r_3$~\cite{jjc2}. Since $\epsilon^r_3$ mainly affects the photon propagation, there is reason to expect that the time information provided by reverberation might significantly improve the constraint. This is indeed confirmed.

As expected, a crucial rule is played by the signal-to-noise, in agreement with what was found in previous studies~\cite{jjc1}. For $N_{\rm line} = 10^3$ counts in the iron line, which can roughly correspond to a good observation of a bright AGN today, the reverberation measurement already provides constraints stronger than the time-integrated iron line, but the CPR deformation parameter $\epsilon^r_3$ is essentially unconstrained, even in the favorable case of a fast-rotating BH observed at a large inclination angle. The time sampling of the reverberation measurement effectively dilutes the signal-to-noise by apportioning the signal into additional time bins.  The measurement is thus significantly affected by Poisson noise of the source, especially where starved for signal. For $N_{\rm line} = 10^4$ line counts, the reverberation measurement is demonstrably superior when compared to the time-integrated constraint. Our simulations show that the CPR deformation parameter $\epsilon^r_3$ can now be constrained, and this is possible even in the most challenging case of a slow-rotating BH observed from a low inclination angle.

The content of the paper is as follows. In Sec.~\ref{s-2}, we briefly review iron line reverberation mapping within the disk-corona model with lamppost geometry. In Sec.~\ref{s-3}, we simulate data and we compare the constraining power of the standard time-integrated iron line measurement with the reverberation one on the CPR deformation parameters $\epsilon^t_3$ and $\epsilon^r_3$. We summarize our results in Sec.~\ref{s-4}. Throughout the paper, we employ units in which $G_{\rm N} = c = 1$ and the convention of a metric with signature $(-+++)$. 

\section{Iron line reverberation mapping \label{s-2}}

In the framework of the corona-disk model with lamppost geometry, the X-ray primary source is point-like and situated along the spin axis of the BH~\cite{lamppost}\footnote{We note that other geometries may also be possible, see e.g.~\cite{sandwich}. The reverberation signal does depend on the geometry of the system. Once we have the time-resolved data, it is straightforward to get the time-integrated signal.}. Such a configuration could be approximately realized, for instance, via a system with a jet wherein the X-ray source would be the jet's compact base. A flash from the X-ray primary source illuminates the cold accretion disk, producing a reflection component which includes emission lines. The iron K$\alpha$ line at $\sim 6.4$\;keV is the most prominent line feature in the resulting X-ray spectrum. Thanks to the finite value of the speed of light, fluoresced emission generated in different regions of the accretion disk reach the distant observer at different times~\cite{reverberation}. The 2D transfer function is the key-quantity that characterize the system: it corresponds to the time evolution of the iron line profile produced by the X-ray source emitting an instantaneous flare.  We neglect any delay due to the atomic reprocessing in the disk, which is expected to be negligibly small for AGN.

In what follows, we employ the usual Novikov-Thorne model~\cite{nt-m} to describe a thin accretion disk surrounding the compact object. In this framework, the disk is in the plane perpendicular to the BH spin, the particles of the gas follow nearly geodesic circular orbits, and the inner edge of the disk is set at the radius of the innermost stable circular orbit (ISCO). The latter assumption plays a crucial role in present spin measurements, and underpins both the disk-continuum and iron-line techniques.  Crucially, the assumption is empirically supported by the observed constancy of the accretion disk inner radius in black hole systems (e.g.,~\cite{lmcx3}).

Our model is specified by the spacetime geometry (spin parameter $a_*$ and possible non-vanishing deformation parameters), the inclination angle of the disk with respect to the line of sight of the distant observer, $i$, and the height of the primary X-ray source, $h$. In all our simulations, the iron line profile is added to a power-law continuum, and therefore we have two additional parameters: the photon index of the continuum, $\Gamma$, and the ratio between the photon number in the iron line and in the continuum. In the lamppost framework, the emissivity profile could be computed self-consistently, e.g., as in Ref.~\cite{dauser}, but here we consider an expedient, instead adopting a simple power-law behavior with a constant emissivity index $q=3$, which corresponds to the Newtonian limit and accordingly must hold at large radii. Since we are not working with real data and are instead illustrating a proof-of-concept via model-based simulations, we expect that this simplification has quite a minor effect on the principal qualitative results.

\section{Simulations in the CPR framework \label{s-3}}

\subsection{Theoretical framework}

In this work we employ the usual approach for testing the Kerr solution, namely we consider a more general background metric characterized by some deformation parameters. The latter are coefficients that parametrize our ignorance and their numerical values are observationally determined. In the Kerr metric, all deformation parameters vanish, and accordingly  the Kerr BH hypothesis is strengthened if the analysis of observational data returns vanishing deformation parameters.

We note that such an approach is very similar to the PPN (Parametrized Post-Newtonian) formalism, which has been successfully used over the last 50 years to test the Schwarzschild solution in the weak gravitational field of the Solar System~\cite{will}. We test the Kerr metric in the same manner the PPN formalism has been used for probing Schwarzschild geometry. It is not possible to directly test the Einstein equations, in the sense that we cannot distinguish a Kerr BH of general relativity from a Kerr BH in another theory of gravity~\cite{kerru}. At the same time, a possible deviation from the Kerr solution does not necessarily imply the breakdown of the Einstein equations, because new physics may arise from, e.g., some exotic matter's energy-momentum tensor, rather than from the gravity sector.

As in Ref.~\cite{jjc2} for the study of the time-integrated iron line profile, we employ the CPR metric~\cite{cpr}. In Boyer-Lindquist coordinates, the line element reads
\begin{widetext}
\be\label{eq-m}
\hspace{-0.8cm}
ds^2 &=& - \left(1 - \frac{2 M r}{\Sigma}\right)\left(1 + h^t\right) dt^2
 - 2 a \sin^2\theta \left[\sqrt{\left(1 + h^t\right)\left(1 + h^r\right)} 
- \left(1 - \frac{2 M r}{\Sigma}\right)\left(1 + h^t\right)\right] dt d\phi \nonumber\\
&& + \frac{\Sigma \left(1 + h^r\right)}{\Delta + h^r a^2 \sin^2\theta} dr^2
+ \Sigma d\theta^2
+ \sin^2\theta \left\{\Sigma + a^2 \sin^2\theta \left[ 2 \sqrt{\left(1 + h^t\right)
\left(1 + h^r\right)} - \left(1 - \frac{2 M r}{\Sigma}\right)
\left(1 + h^t\right)\right]\right\} d\phi^2 \, , 
\ee
\end{widetext}
where $\Sigma = r^2 + a^2 \cos^2 \theta$, $\Delta = r^2 - 2 M r + a^2$, and
\be
h^t &=& \sum_{k=0}^{+\infty} \left(\epsilon_{2k}^t 
+ \epsilon_{2k+1}^t \frac{M r}{\Sigma}\right)\left(\frac{M^2}{\Sigma}
\right)^k\, , \\
h^r &=& \sum_{k=0}^{+\infty} \left(\epsilon_{2k}^r
+ \epsilon_{2k+1}^r \frac{M r}{\Sigma}\right)\left(\frac{M^2}{\Sigma}
\right)^k \, .
\ee
There are two infinite sets of deformation parameters, $\{\epsilon_k^t\}$ and $\{\epsilon_k^r\}$ (at increasingly high order). Since the lowest order deformation parameters are already strongly constrained to recover the Newtonian limit and meet the PPN bounds (see~\cite{cpr} for more details), in what follows we consider the leading plausible deformation parameters $\epsilon_3^t$ and $\epsilon_3^r$. Higher order deformation parameters have diminishing effect and are accordingly omitted for the sake of simplicity.

\begin{figure*}
\begin{center}
\includegraphics[type=pdf,ext=.pdf,read=.pdf,width=7.8cm]{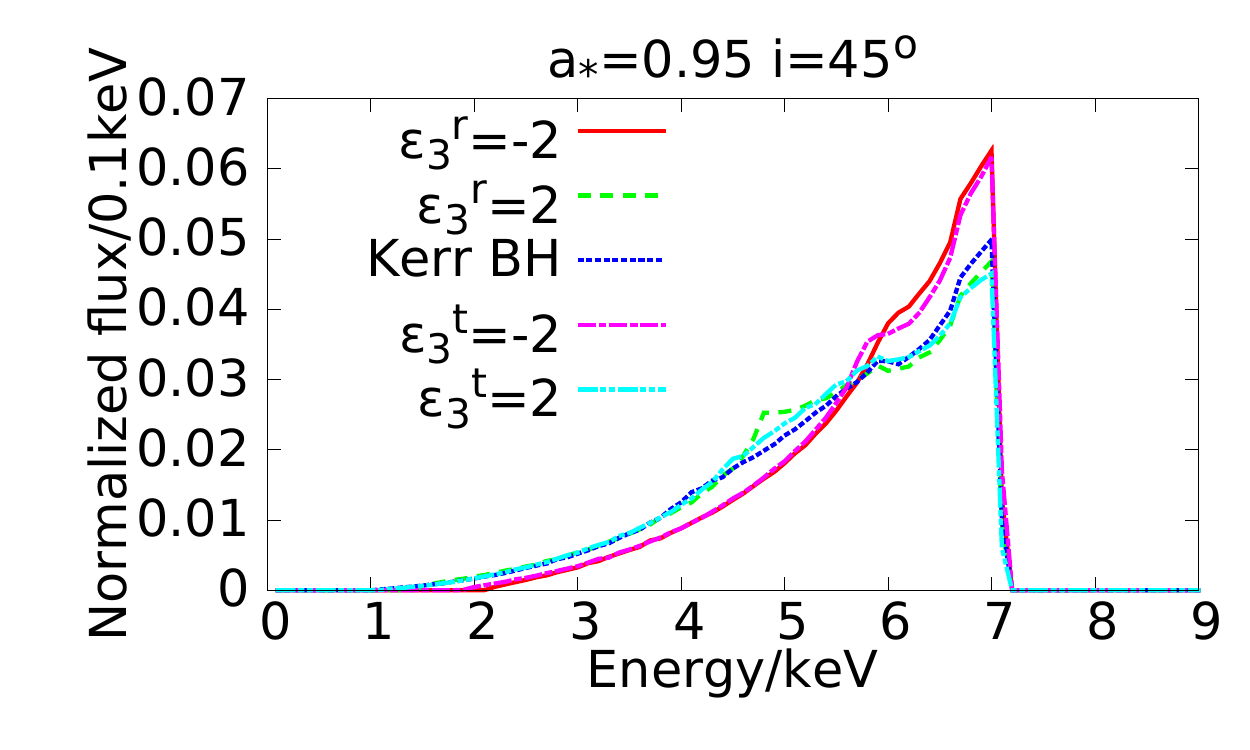}
\hspace{0.5cm}
\includegraphics[type=pdf,ext=.pdf,read=.pdf,width=7.8cm]{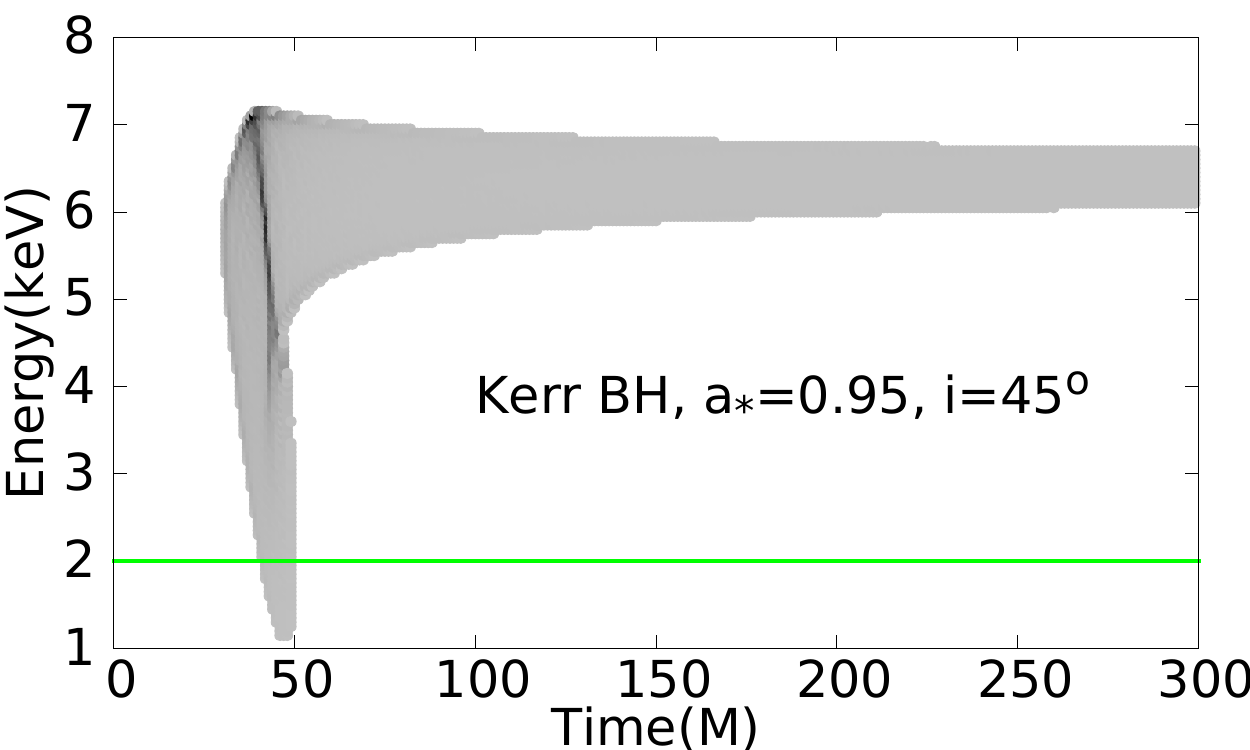}\\ 
\vspace{0.8cm}
\includegraphics[type=pdf,ext=.pdf,read=.pdf,width=7.8cm]{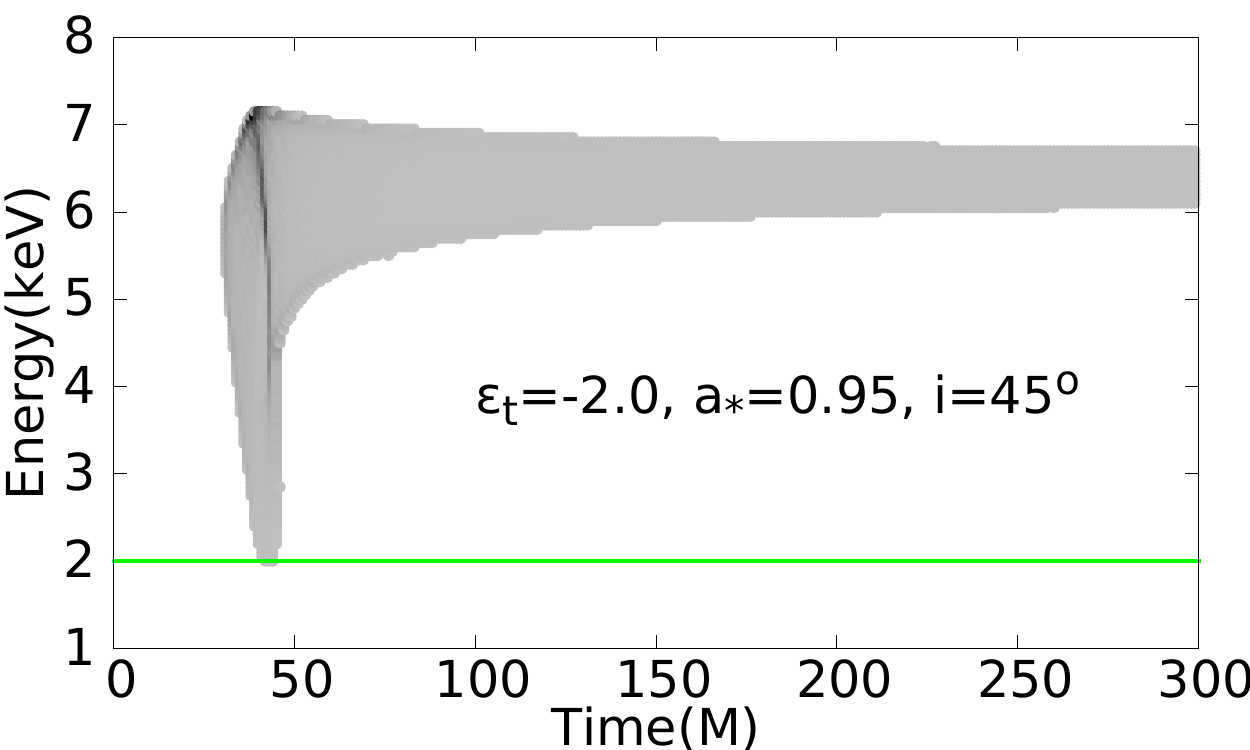}
\hspace{0.5cm}
\includegraphics[type=pdf,ext=.pdf,read=.pdf,width=7.8cm]{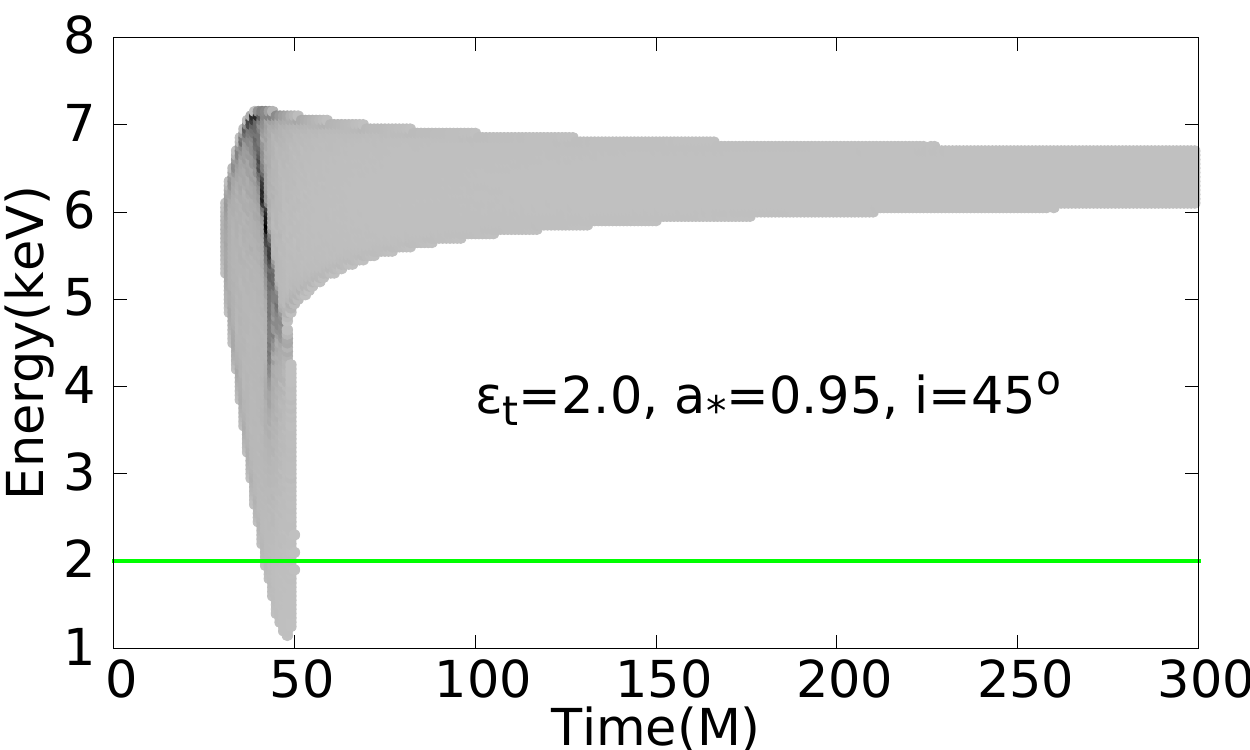}\\
\vspace{0.8cm}
\includegraphics[type=pdf,ext=.pdf,read=.pdf,width=7.8cm]{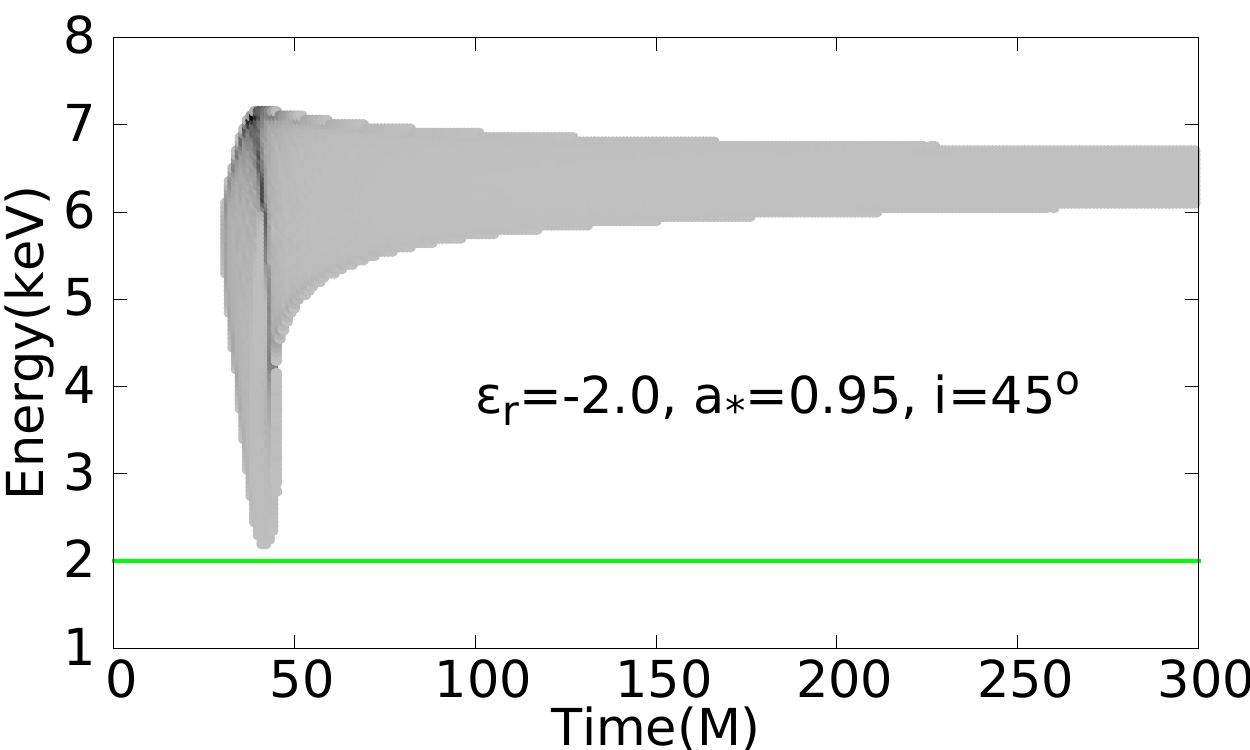}
\hspace{0.5cm}
\includegraphics[type=pdf,ext=.pdf,read=.pdf,width=7.8cm]{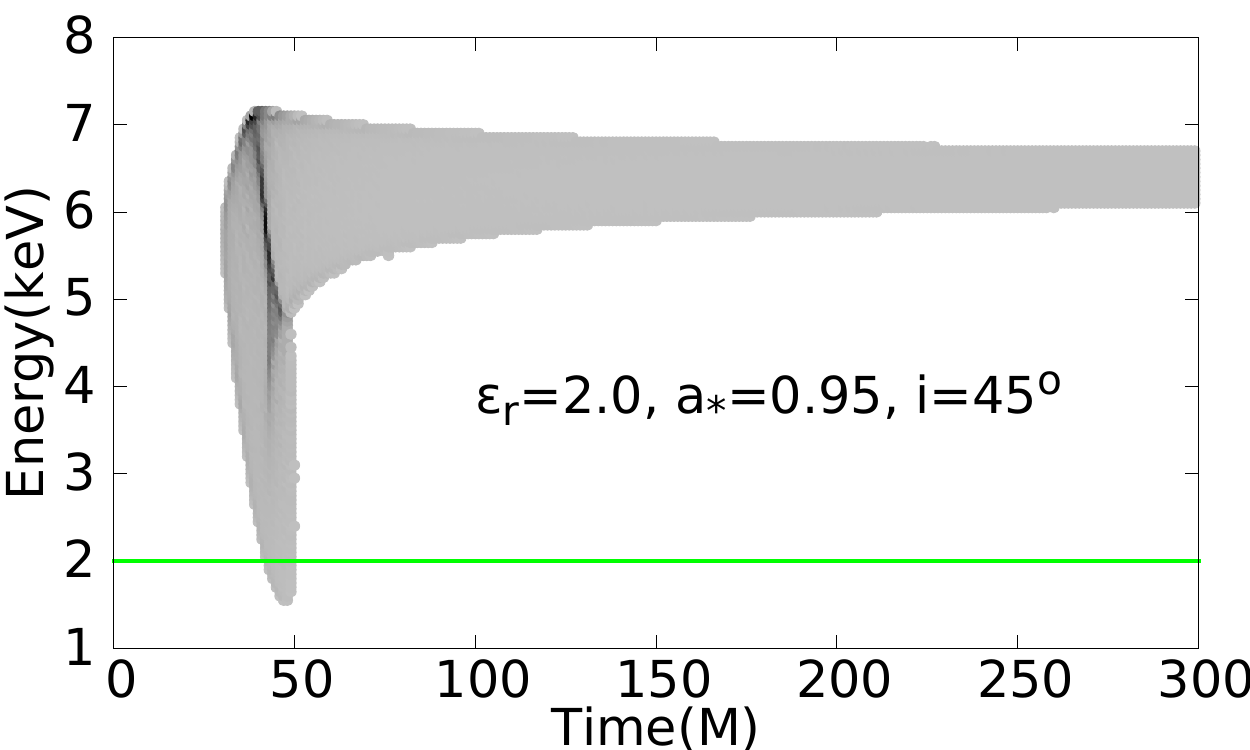}
\end{center}
\caption{Impact of the deformation parameters $\epsilon^t_3$ and $\epsilon^r_3$ on the time-integrated iron line profile (top left panel) and on the 2D transfer function: top right panel for $\epsilon^t_3 = \epsilon^r_3 = 0$ (Kerr black hole), central left panel for $\epsilon^t_3 = -2$ and $\epsilon^r_3 = 0$, central right panel for $\epsilon^t_3 = 2$ and $\epsilon^r_3 = 0$, bottom left panel for $\epsilon^t_3 = 0$ and $\epsilon^r_3 = -2$, bottom right panel for $\epsilon^t_3 = 0$ and $\epsilon^r_3 = 2$. The spin parameter is $a_* = 0.95$ and the inclination angle is $i = 45^\circ$. In the plots of the 2D transfer function, the color indicates the photon number density and ranges from light gray to black as the density increases (in arbitrary units).}
\label{fig0}
\end{figure*}

\subsection{Results}

To assess the constraining capability of an iron line reverberation measurement and compare it with that of the standard time-integrated (spectral) iron line observation, we proceed as in Refs.~\cite{jjc1,jjc2}. We simulate data using a reference model, namely a Kerr BH with spin parameter $a_*'$ and observed from an inclination angle $i'$. As stated above, for simplicity the emissivity profile is taken to be a power law, $I_{\rm e} \propto r^{-3}$. We compute both the time-integrated iron line profile and the 2D transfer function, assuming an energy resolution $\Delta E = 50$~eV and in the latter case a time resolution $\Delta t = M$ (for instance, if $M = 10^6$~$M_\odot$, $\Delta t \approx 5$~s). In the case of the reverberation measurement, the height of the source is a free parameter to be determined by the fit. In our reference model, the height of the source is $h' = 10$~$M$. Concerning the power-law continuum, in the reference model it is normalized to include 100~times the number of iron line photons when integrated over the energy range 1--9~keV, and adopting a nominal photon index $\Gamma' = 2$. This choice corresponds to an equivalent width of $EW \approx 400$~eV, with the precise value depending weakly on the line shape. In the analysis below, the ratio between the continuum and the iron line photon flux, $K$, as well as the photon index of the continuum, $\Gamma$, are left as free parameters to be determined by the fit. The impact of the deformation parameter $\epsilon^t_3$ and $\epsilon^r_3$ on both the time-integrated and the reverberation measurements is illustrated in Fig.~\ref{fig0} (where the power-law continuum has been removed to better visualize the effects of the deformation parameters).

We treat these simulated spectra as real data; they include Poisson noise, and the spectra are binned to achieve a minimum count-number per bin  ($n_{\rm min} = 20$). These spectra are fitted using a BH model that allows for CPR deformation parameters $\epsilon^t_3$ and $\epsilon^r_3$, viewing angle $i$, height of the source $h$, photon index of the continuum $\Gamma$, the ratio between the continuum and the iron line photon flux $K$, and black hole spin $a_*$. We fit using standard $\chi^2$ analysis and use contour levels of $\Delta\chi^2  = \chi^2 - \chi_{min}^2$ to discern the strength of constraints that can be obtained.

\begin{figure*}
\begin{center}
\includegraphics[type=pdf,ext=.pdf,read=.pdf,width=7.0cm]{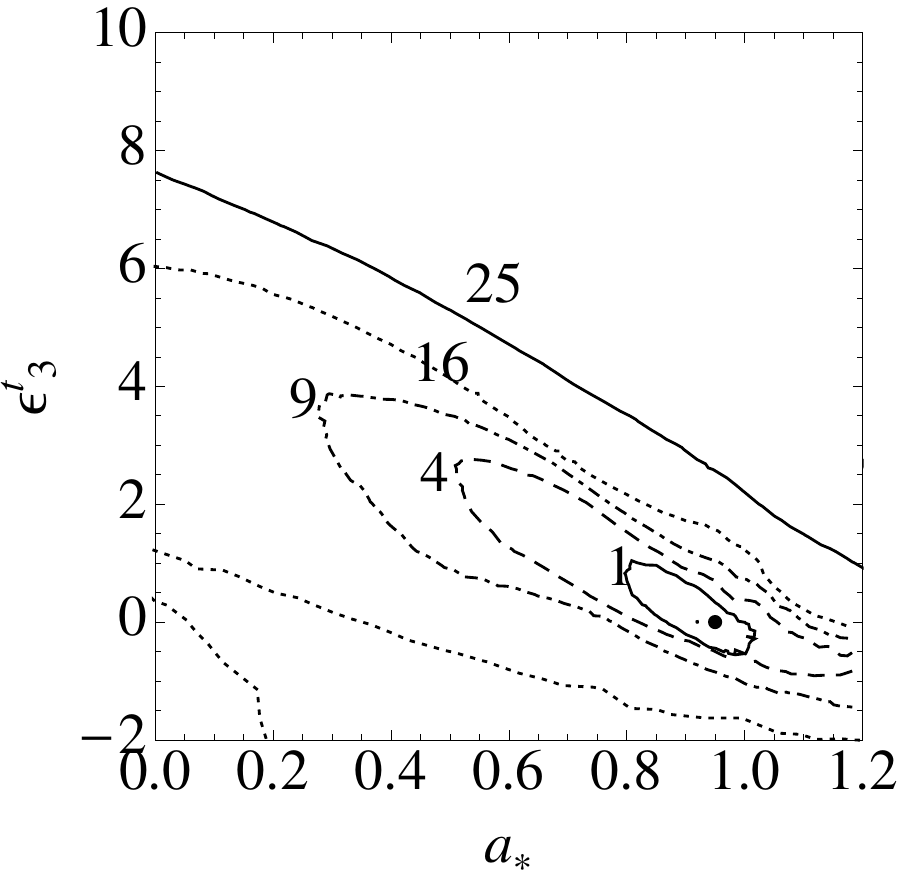}
\hspace{0.8cm}
\includegraphics[type=pdf,ext=.pdf,read=.pdf,width=7.0cm]{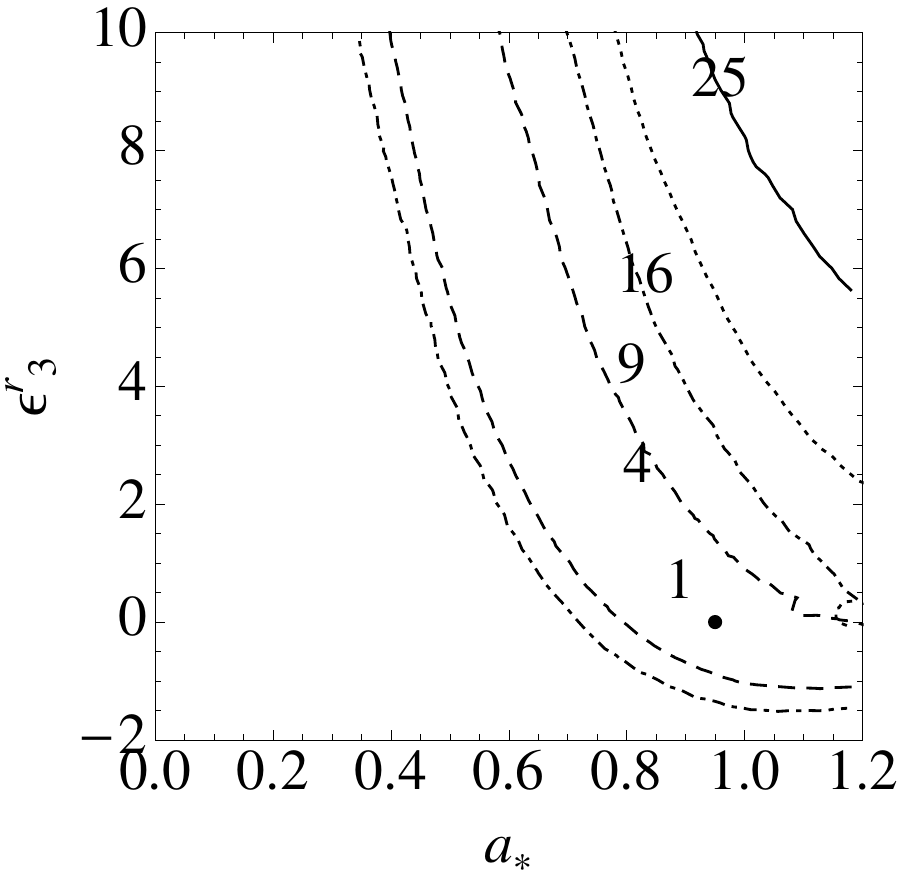}\\ 
\vspace{0.8cm}
\includegraphics[type=pdf,ext=.pdf,read=.pdf,width=7.0cm]{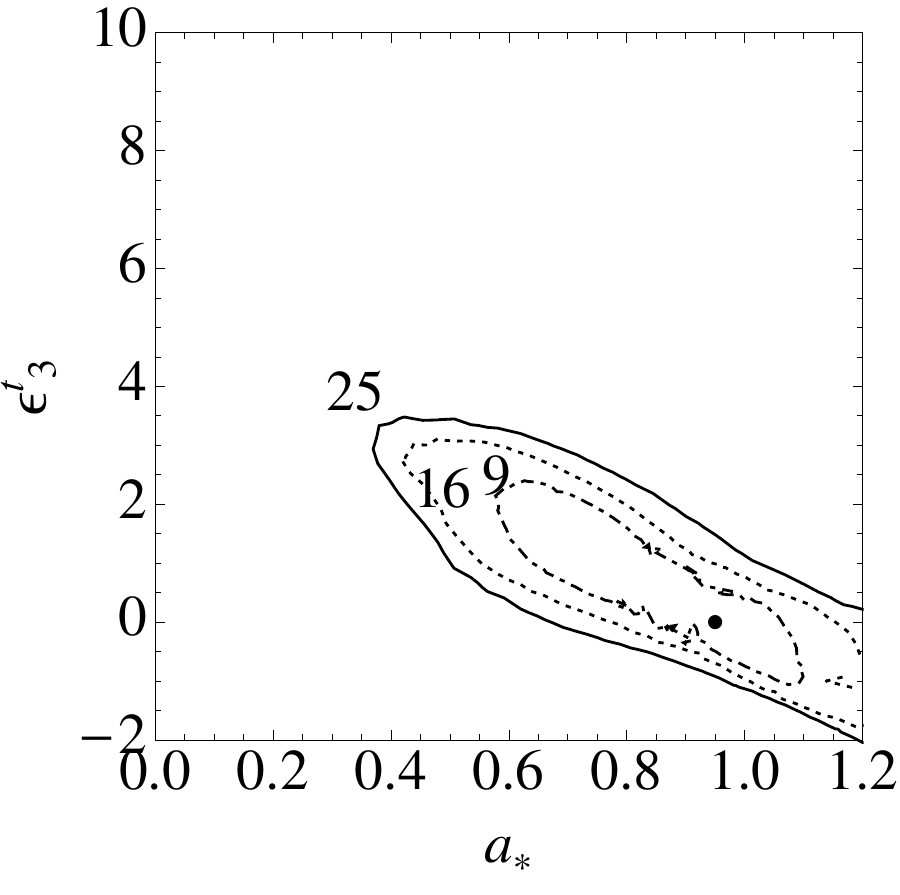}
\hspace{0.8cm}
\includegraphics[type=pdf,ext=.pdf,read=.pdf,width=7.0cm]{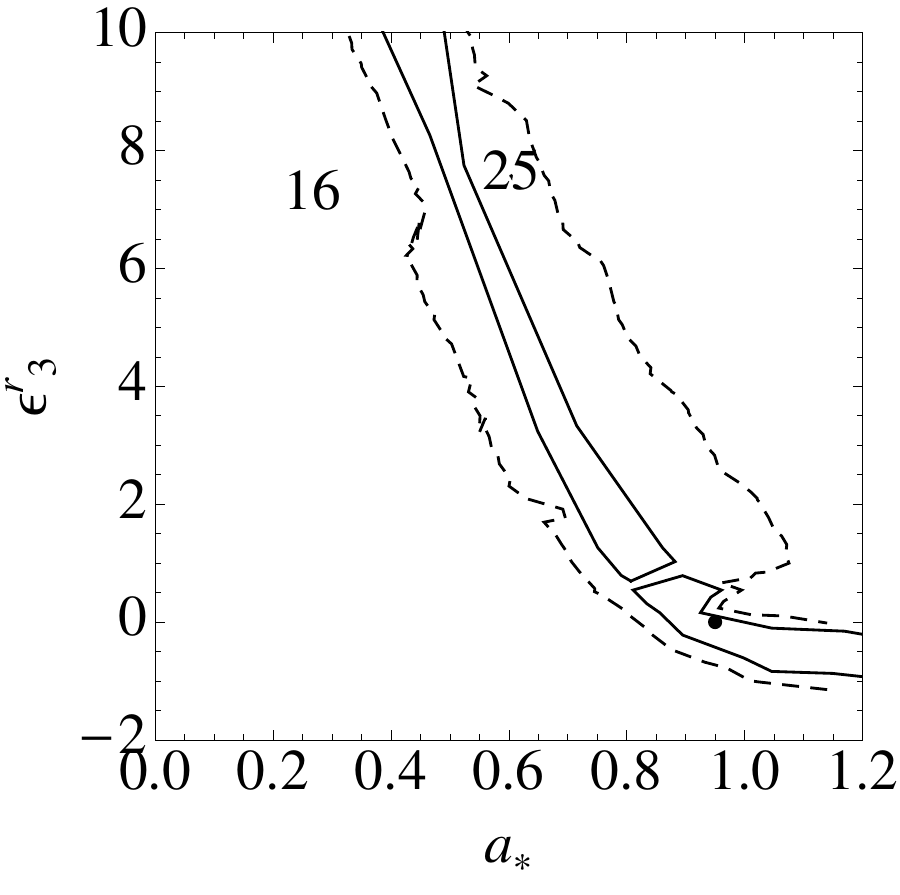}
\end{center}
\caption{$\Delta\chi^2$ contours with $N_{\rm line} = 10^3$ photons (top panels) and $N_{\rm line} = 10^4$ (bottom panels) from the analysis of the iron line profile. The data were simulated from a reference model consisting of a Kerr BH with spin parameter $a_*' = 0.95$ and inclination angle $i' = 70^\circ$. We fit for spin and inclination, and in the left panels we allow for a non-vanishing $\epsilon^t_3$ while we assume $\epsilon^r_3 = 0$. In the right panels we illustrate the converse measurement, namely $\epsilon^t_3 = 0$ while $\epsilon^r_3$ can vary. The ratio between the continuum and the iron line photon flux, $K$, as well as the photon index of the continuum, $\Gamma$, are also free parameters in the fit. }
\label{fig1}
\end{figure*}

\begin{figure*}
\begin{center}
\includegraphics[type=pdf,ext=.pdf,read=.pdf,width=7.0cm]{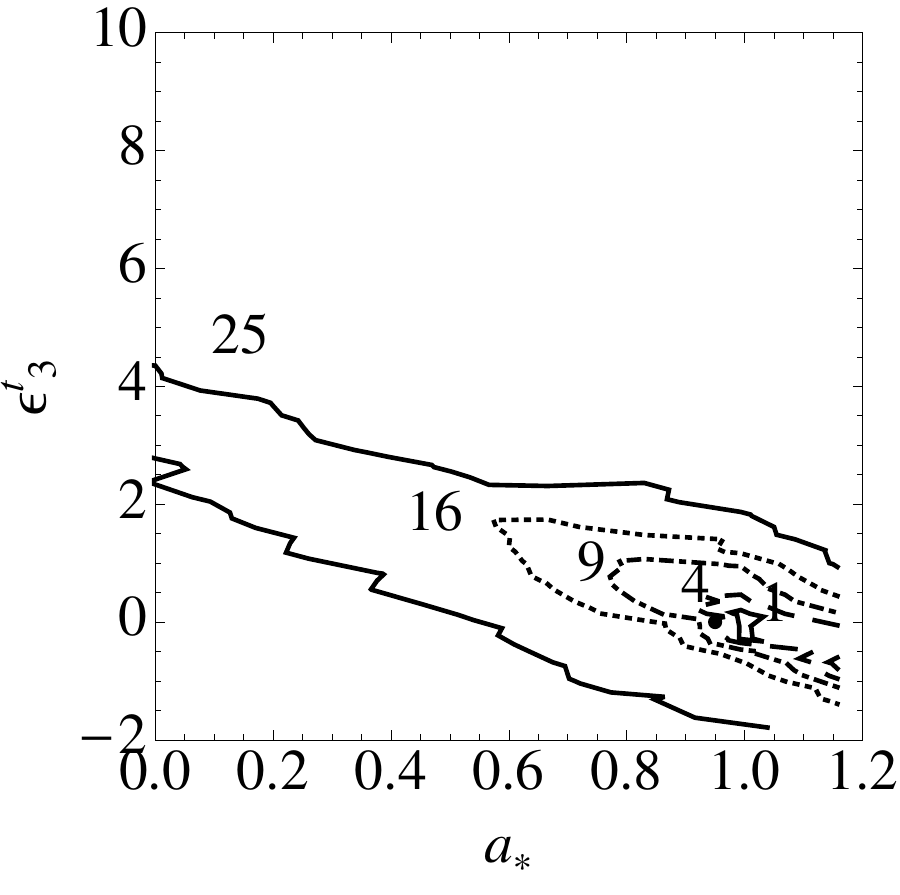}
\hspace{0.8cm}
\includegraphics[type=pdf,ext=.pdf,read=.pdf,width=7.0cm]{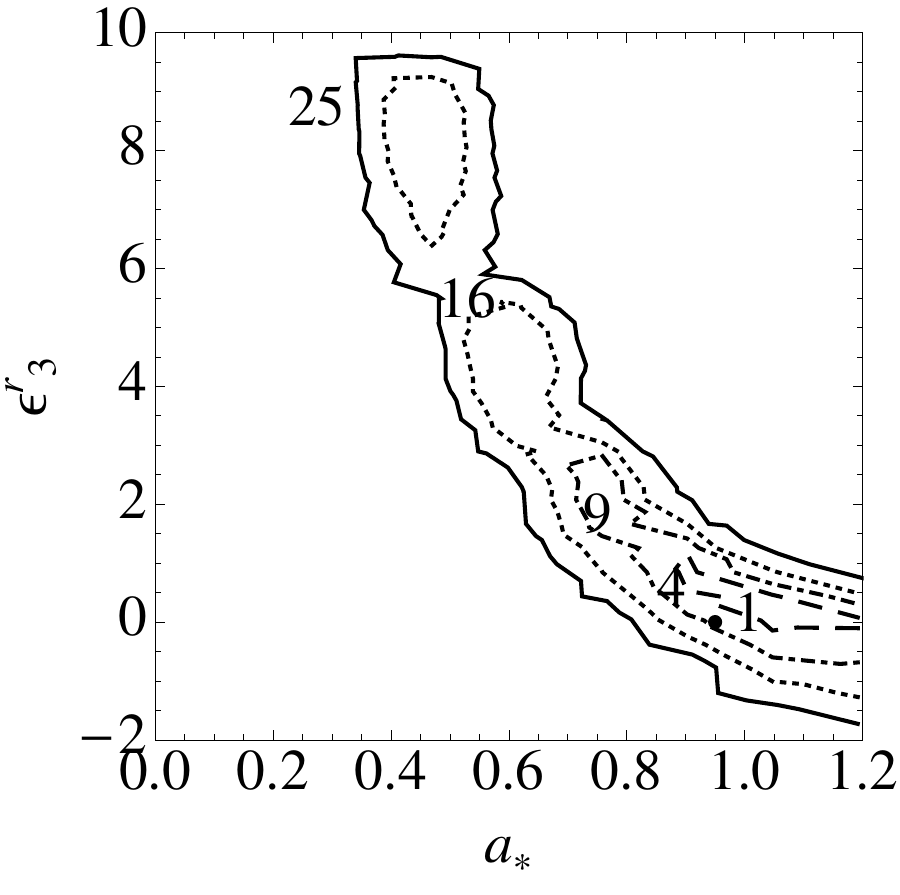}\\ 
\vspace{0.8cm}
\includegraphics[type=pdf,ext=.pdf,read=.pdf,width=7.0cm]{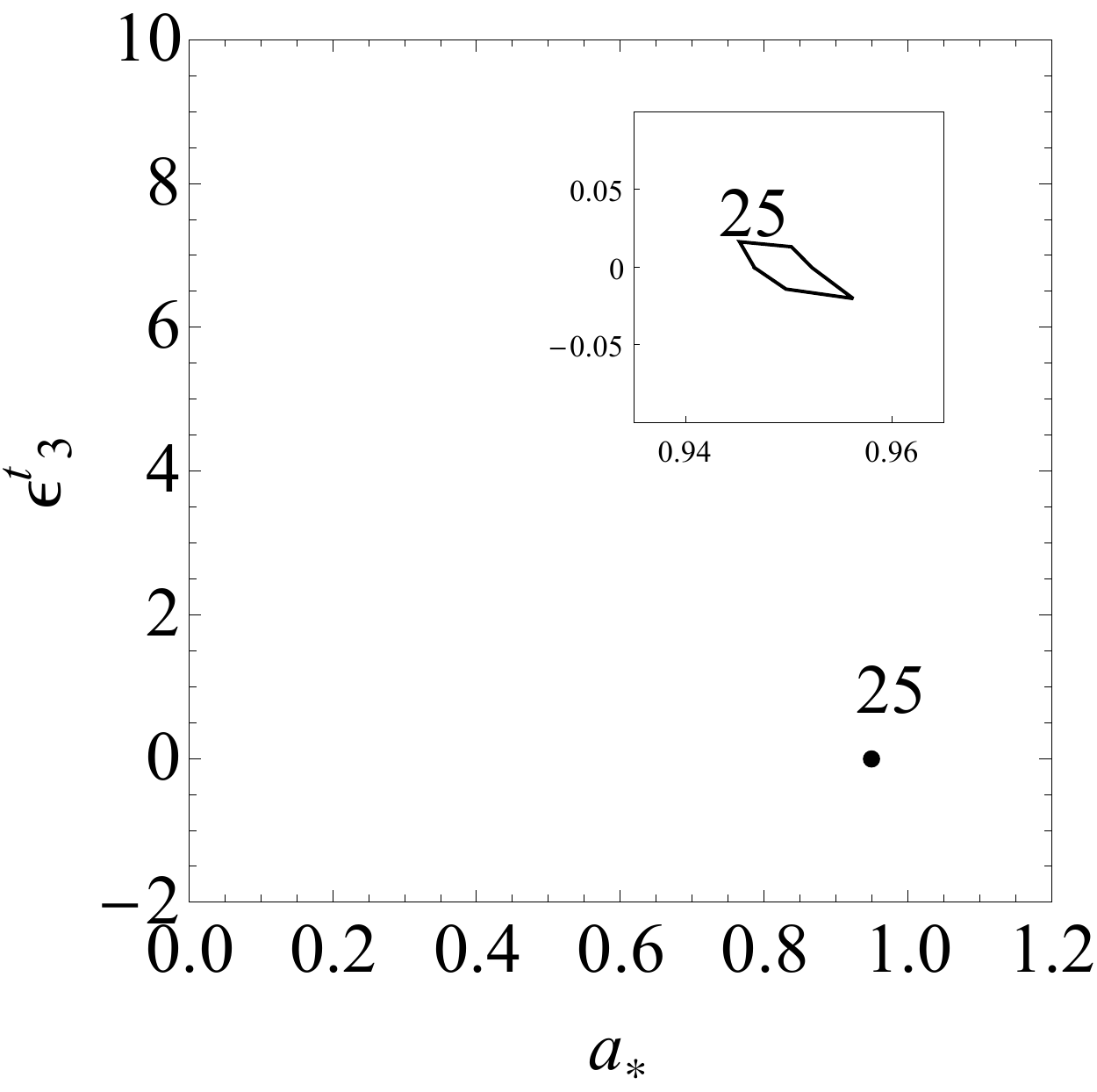}
\hspace{0.8cm}
\includegraphics[type=pdf,ext=.pdf,read=.pdf,width=7.0cm]{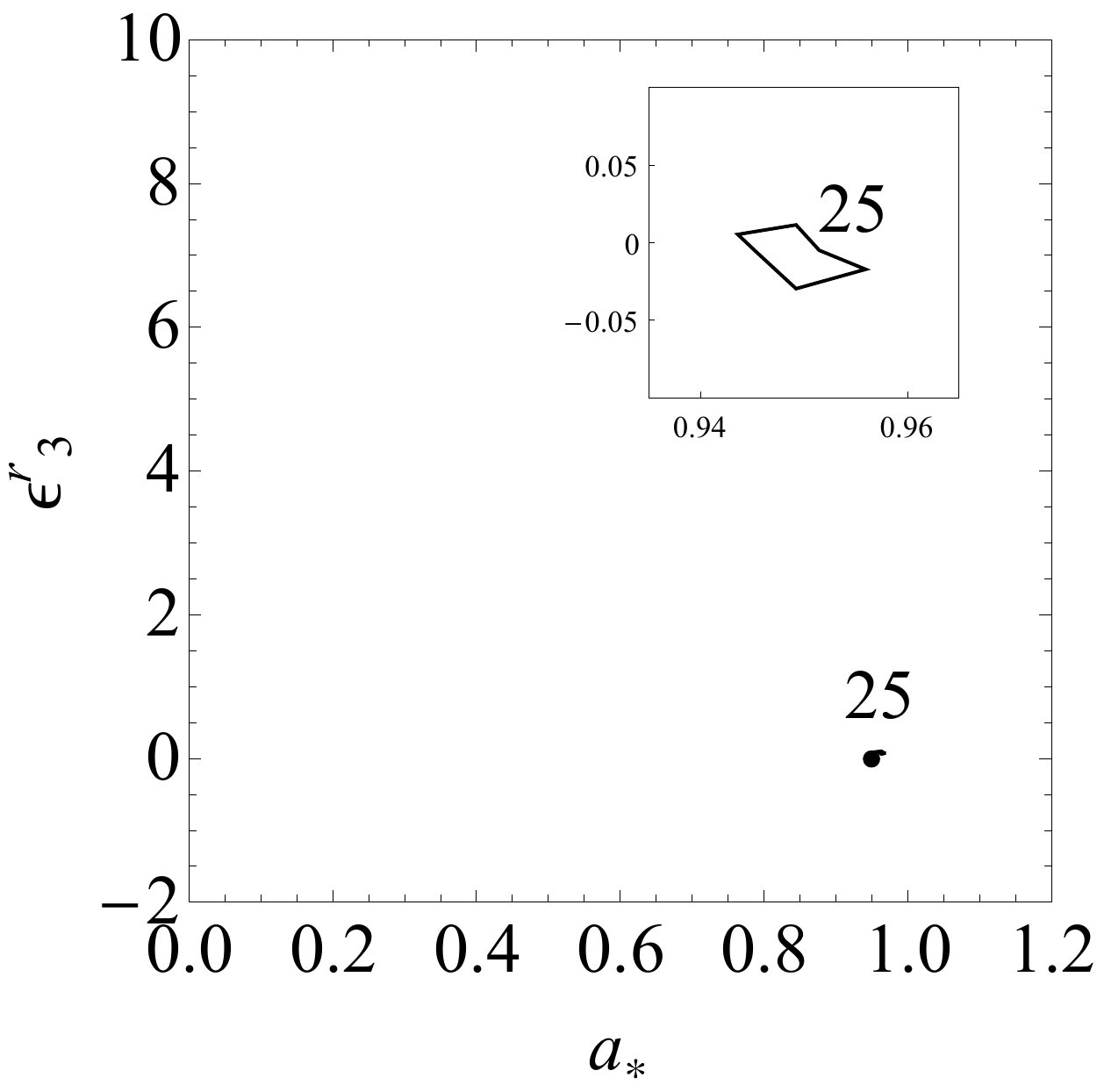}
\end{center}
\caption{Compare to Fig.~\ref{fig1}.  $\Delta\chi^2$ contours with $N_{\rm line} = 10^3$ photons (top panels) and $N_{\rm line} = 10^4$ (bottom panels) from the analysis of the 2D transfer function. The reference model is a Kerr BH with spin parameter $a_*' = 0.95$ and inclination angle $i' = 70^\circ$. In the left panels, we allow for a non-vanishing $\epsilon^t_3$ and we assume $\epsilon^r_3 = 0$. In the right panels we consider the converse case, namely $\epsilon^t_3 = 0$ while $\epsilon^r_3$ can vary. The height of the source $h$, the ratio between the continuum and the iron line photon flux, $K$, and the photon index of the continuum, $\Gamma$, are also left as fit parameters. See the text for more details.}
\label{fig2}
\end{figure*}

\begin{figure*}
\begin{center}
\includegraphics[type=pdf,ext=.pdf,read=.pdf,width=7.0cm]{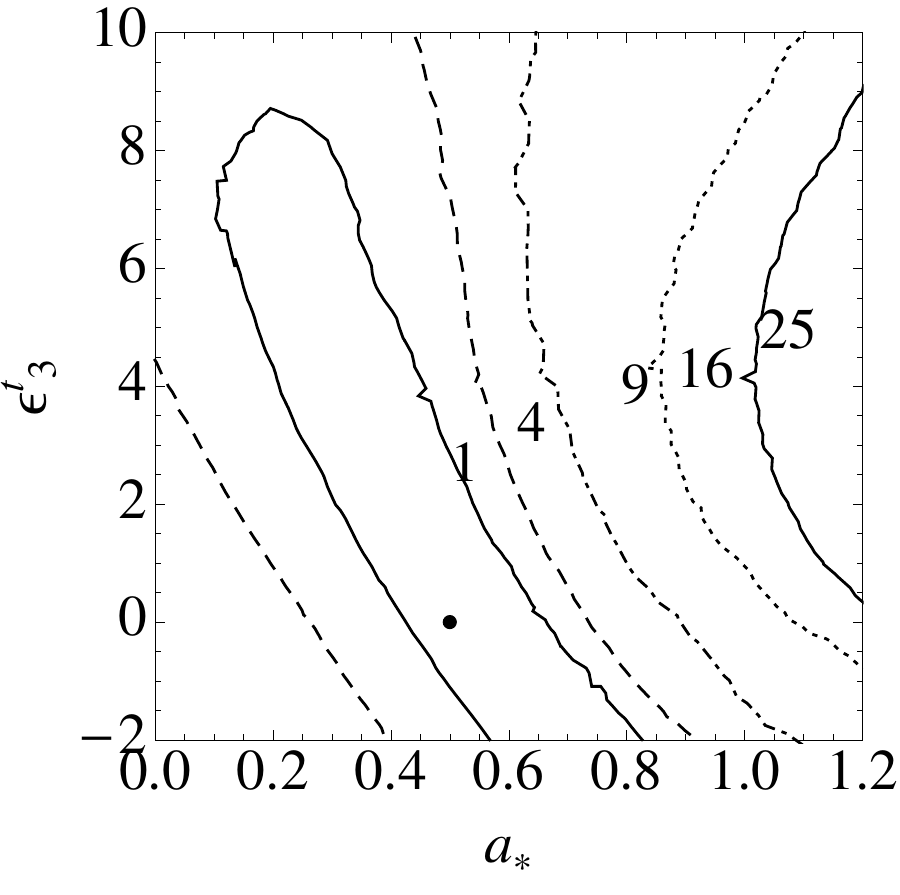}
\hspace{0.8cm}
\includegraphics[type=pdf,ext=.pdf,read=.pdf,width=7.0cm]{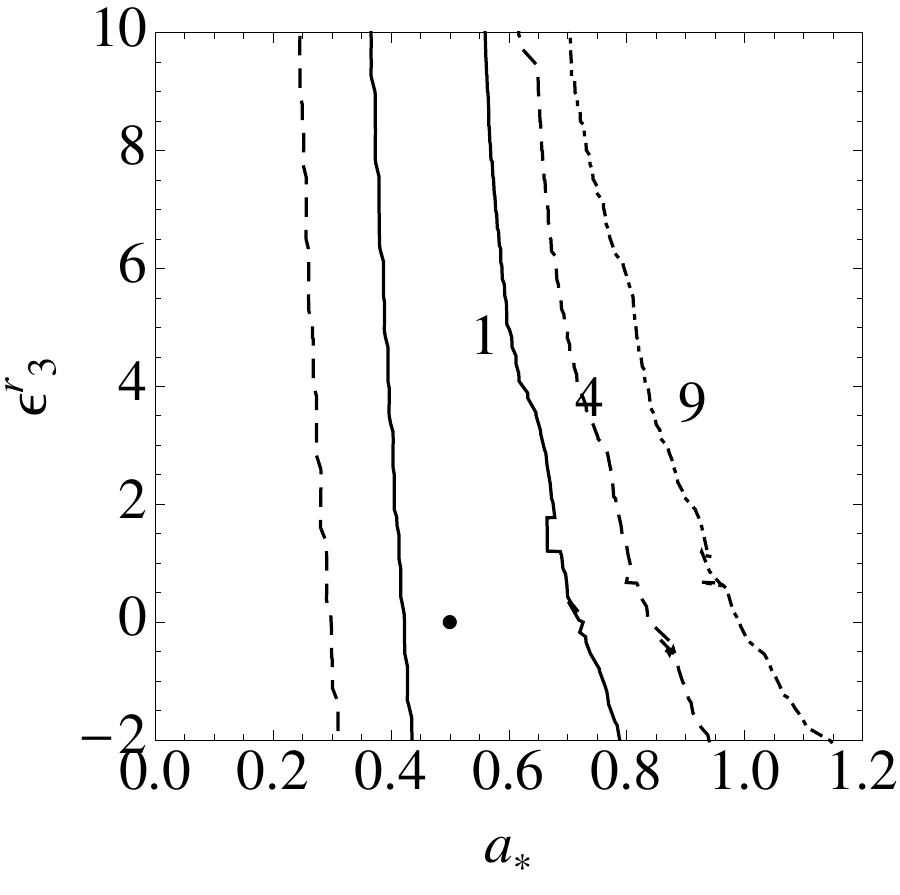}\\ 
\vspace{0.8cm}
\includegraphics[type=pdf,ext=.pdf,read=.pdf,width=7.0cm]{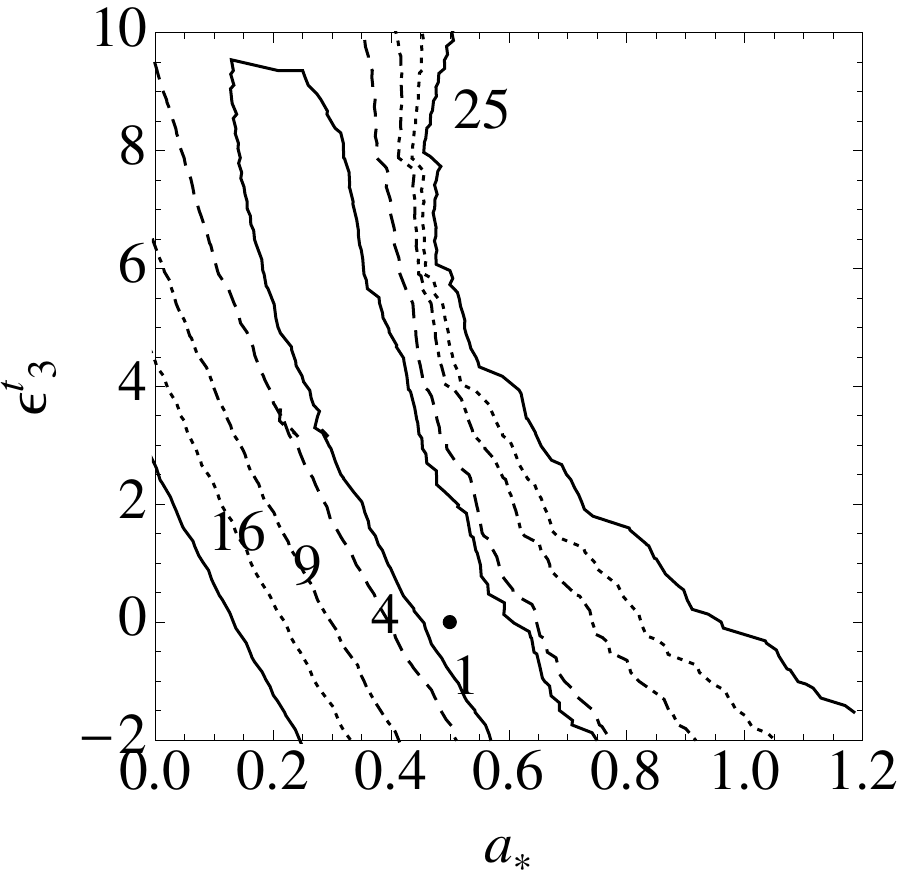}
\hspace{0.8cm}
\includegraphics[type=pdf,ext=.pdf,read=.pdf,width=7.0cm]{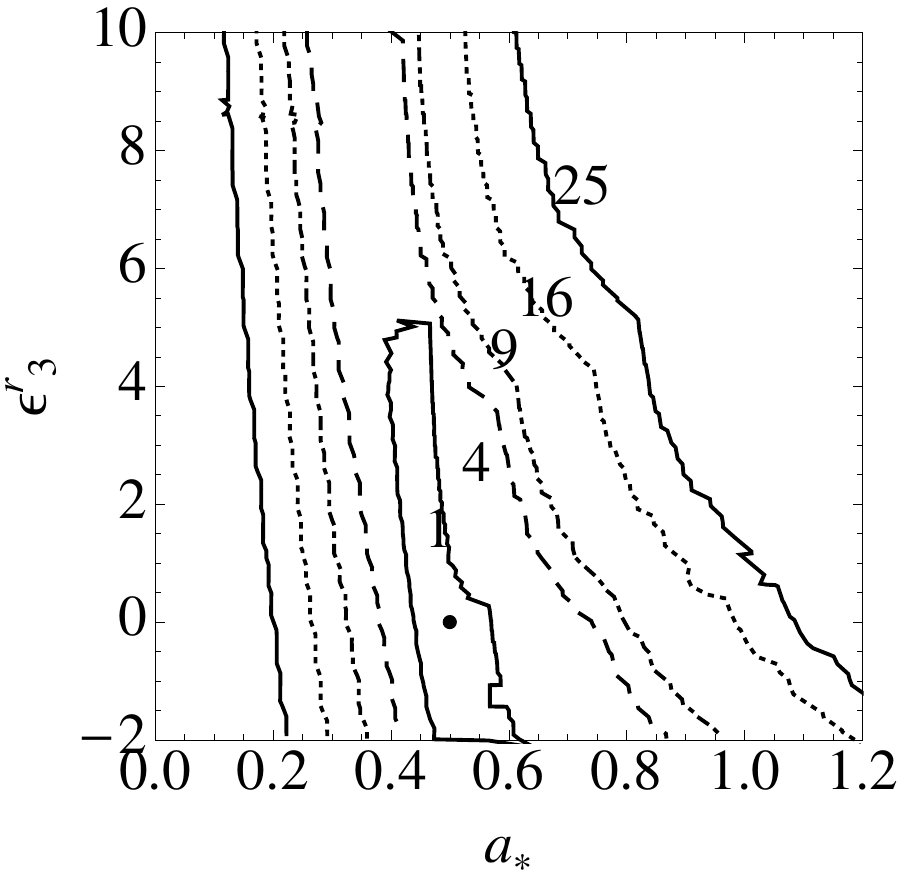}
\end{center}
\caption{$\Delta\chi^2$ contours with $N_{\rm line} = 10^3$ photons (top panels) and $N_{\rm line} = 10^4$ (bottom panels) from the analysis of the iron line profile. The reference model is a Kerr BH with spin parameter $a_*' = 0.5$ and inclination angle $i' = 20^\circ$. In the left panels, we allow for a non-vanishing $\epsilon^t_3$ and we assume $\epsilon^r_3 = 0$. In the right panels we consider the converse case, namely $\epsilon^t_3 = 0$ while $\epsilon^r_3$ can vary. The ratio between the continuum and the iron line photon flux, $K$, as well as the photon index of the continuum, $\Gamma$, are also free parameters in the fit.}
\label{fig3}
\end{figure*}

\begin{figure*}
\begin{center}
\includegraphics[type=pdf,ext=.pdf,read=.pdf,width=7.0cm]{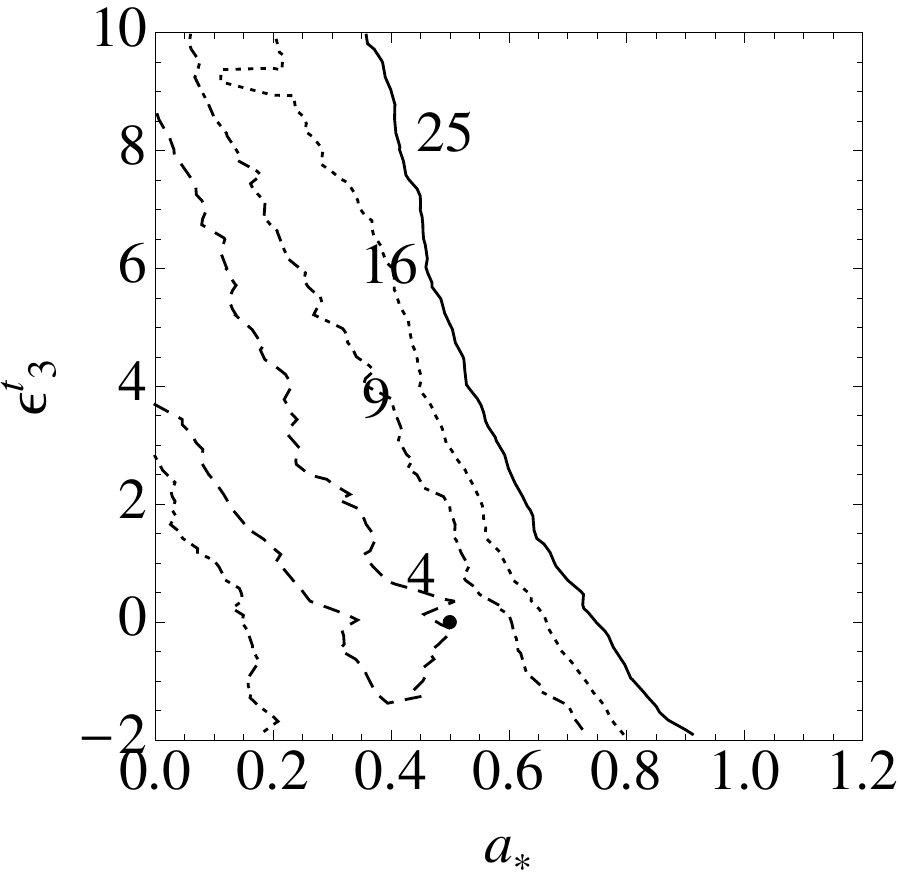}
\hspace{0.8cm}
\includegraphics[type=pdf,ext=.pdf,read=.pdf,width=7.0cm]{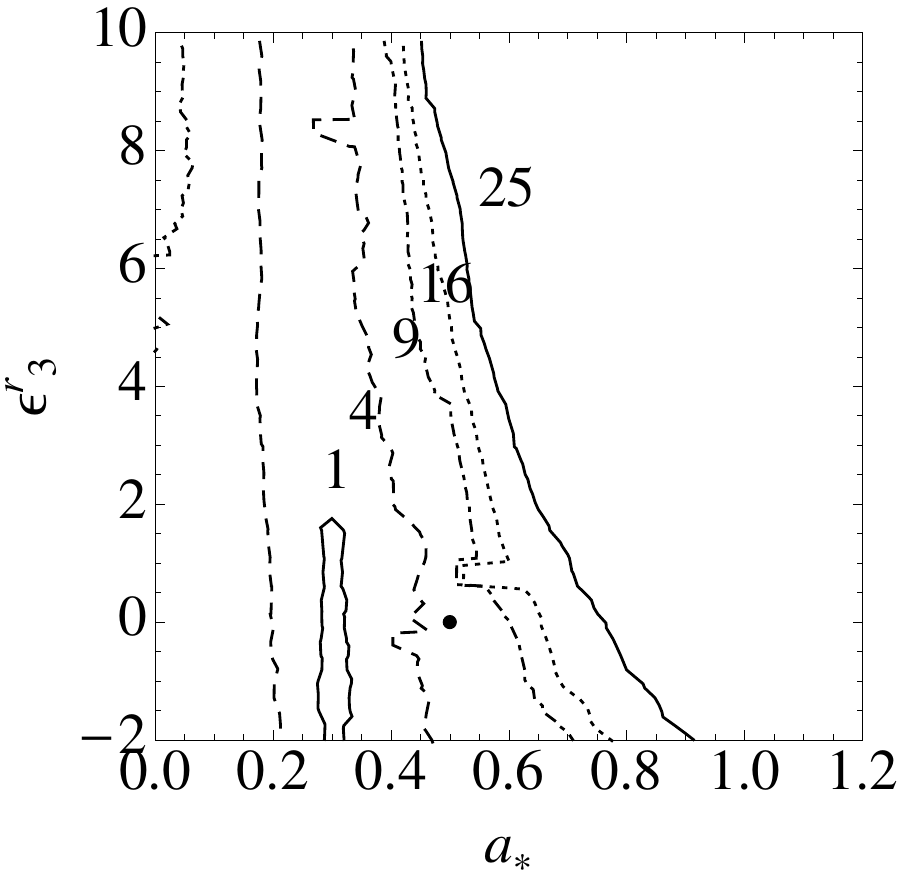}\\ 
\vspace{0.8cm}
\includegraphics[type=pdf,ext=.pdf,read=.pdf,width=7.0cm]{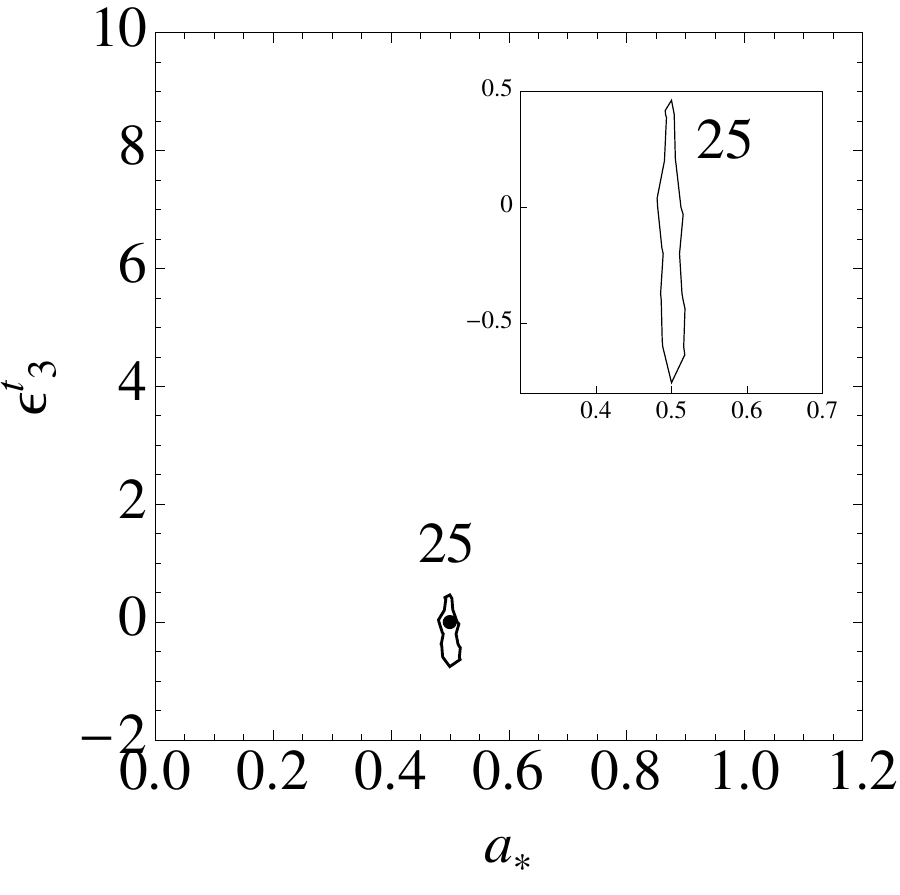}
\hspace{0.8cm}
\includegraphics[type=pdf,ext=.pdf,read=.pdf,width=7.0cm]{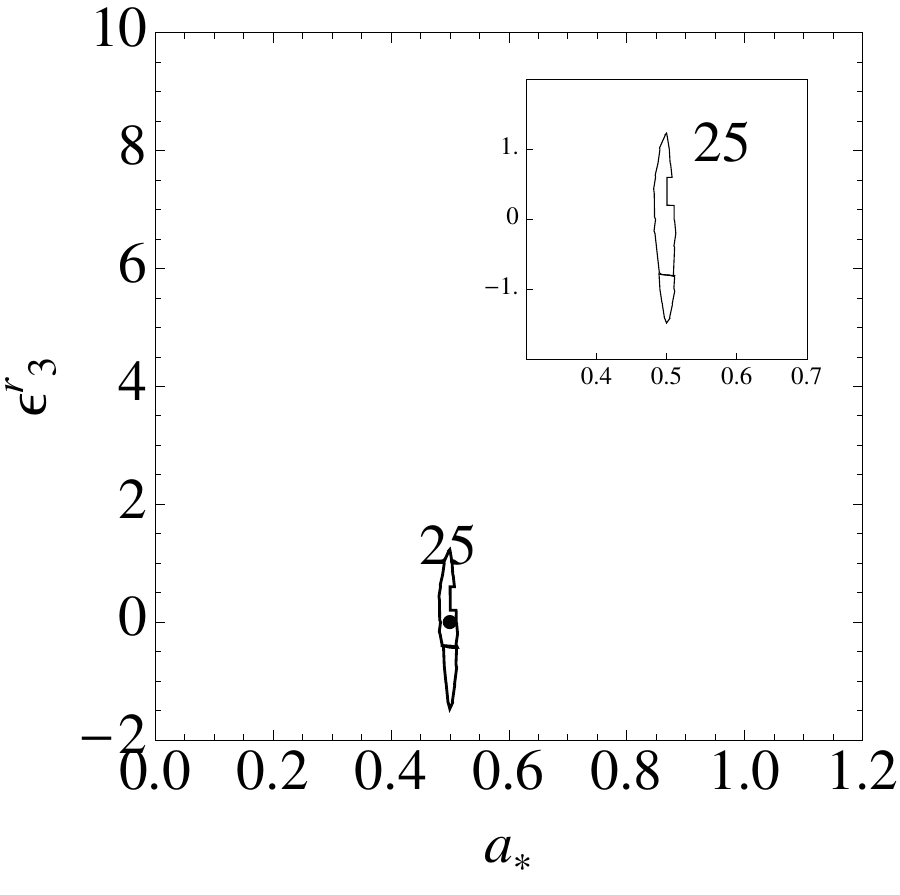}
\end{center}
\caption{Compare to Fig.~\ref{fig3}.  $\Delta\chi^2$ contours with $N_{\rm line} = 10^3$ photons (top panels) and $N_{\rm line} = 10^4$ (bottom panels) from the analysis of the 2D transfer function. The reference model is a Kerr BH with spin parameter $a_*' = 0.5$ and inclination angle $i' = 20^\circ$. In the left panels, we allow for a non-vanishing $\epsilon^t_3$ and we assume $\epsilon^r_3 = 0$. In the right panels we consider the converse case, namely $\epsilon^t_3 = 0$ while $\epsilon^r_3$ can vary. The height of the source $h$, the ratio between the continuum and the iron line photon flux, $K$, and the photon index of the continuum, $\Gamma$, are also left as fit parameters. See the text for more details.}
\label{fig4}
\end{figure*}

\begin{figure*}
\begin{center}
\includegraphics[type=pdf,ext=.pdf,read=.pdf,width=7.0cm]{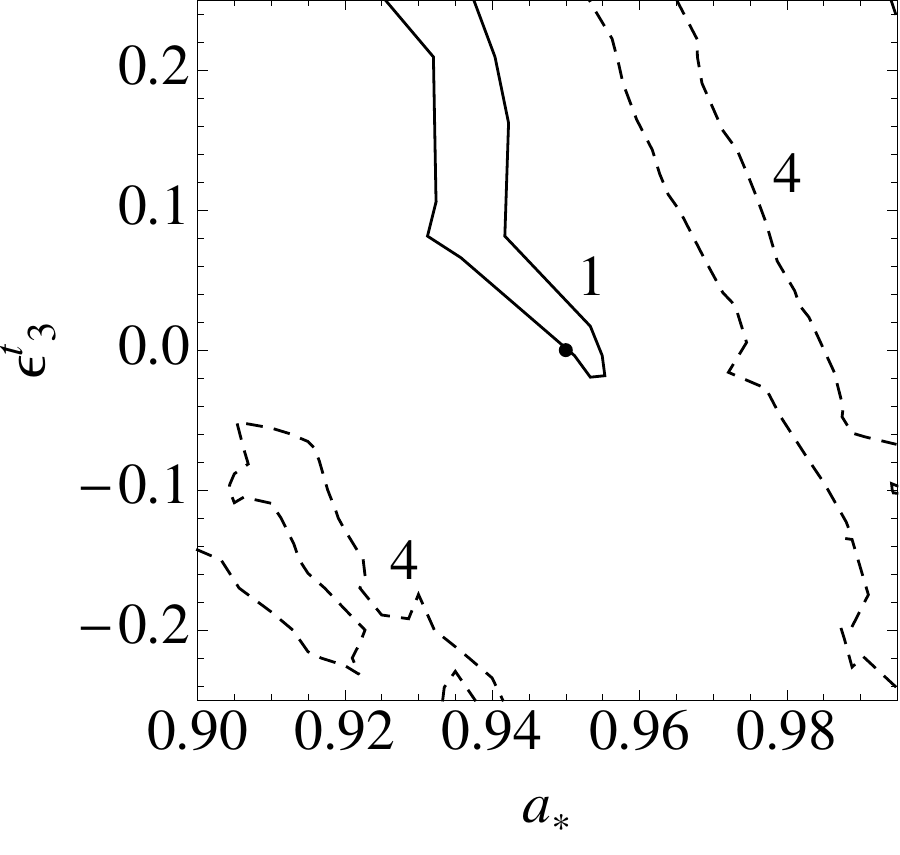}
\hspace{0.8cm}
\includegraphics[type=pdf,ext=.pdf,read=.pdf,width=7.0cm]{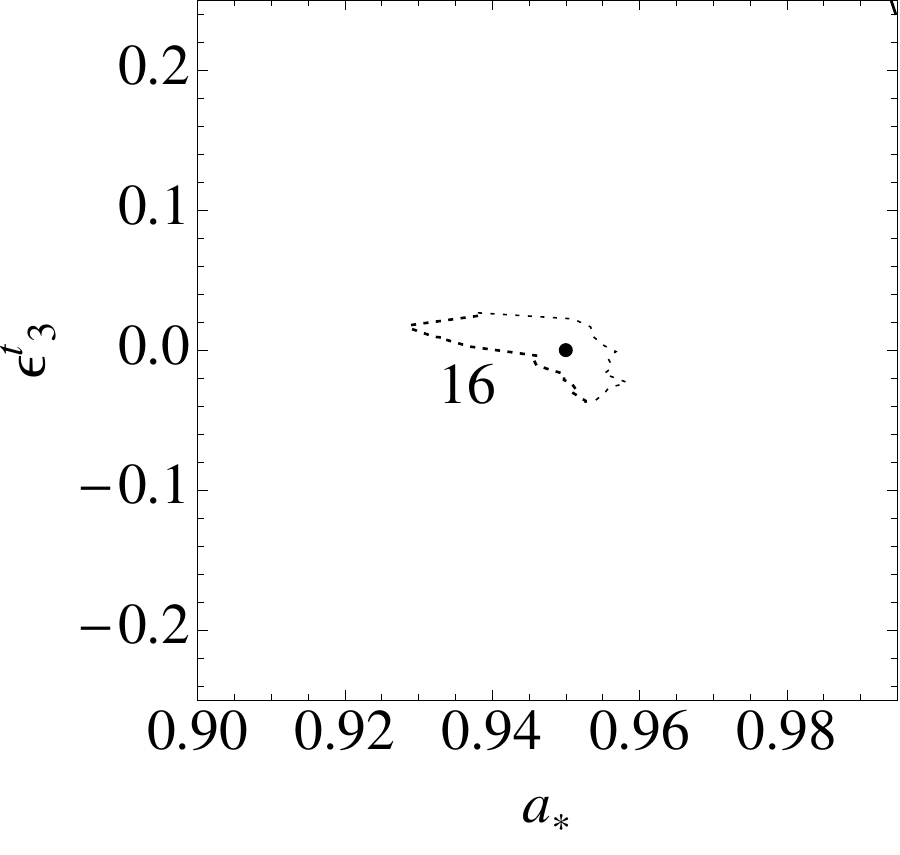}\\ 
\vspace{0.8cm}
\includegraphics[type=pdf,ext=.pdf,read=.pdf,width=7.0cm]{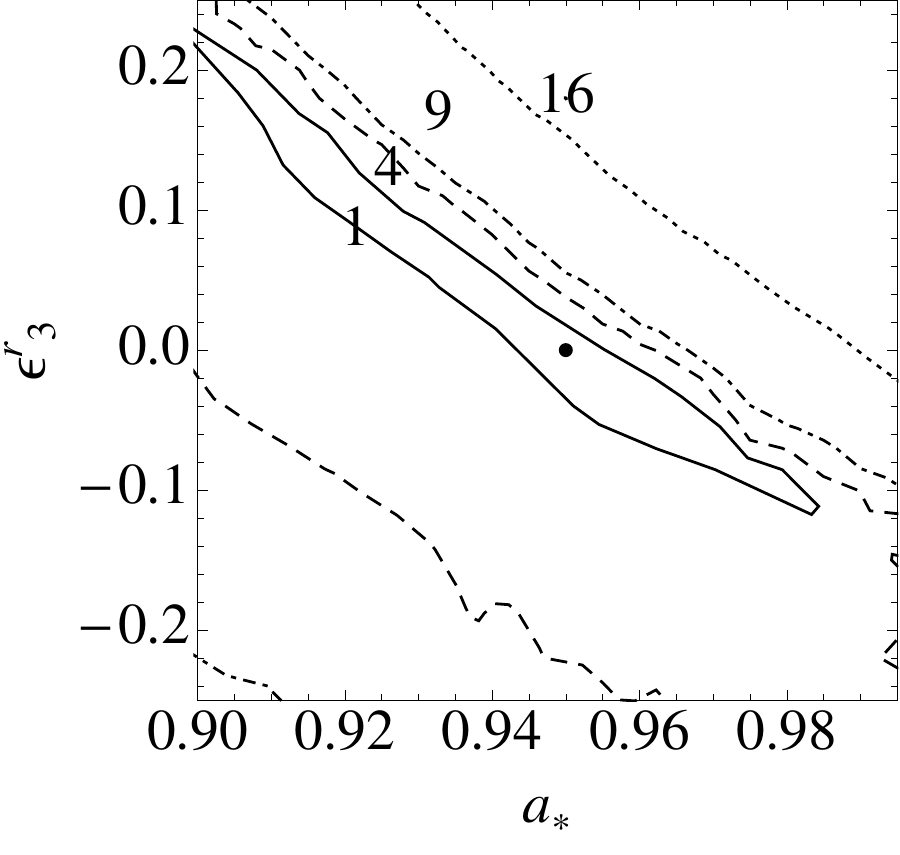}
\hspace{0.8cm}
\includegraphics[type=pdf,ext=.pdf,read=.pdf,width=7.0cm]{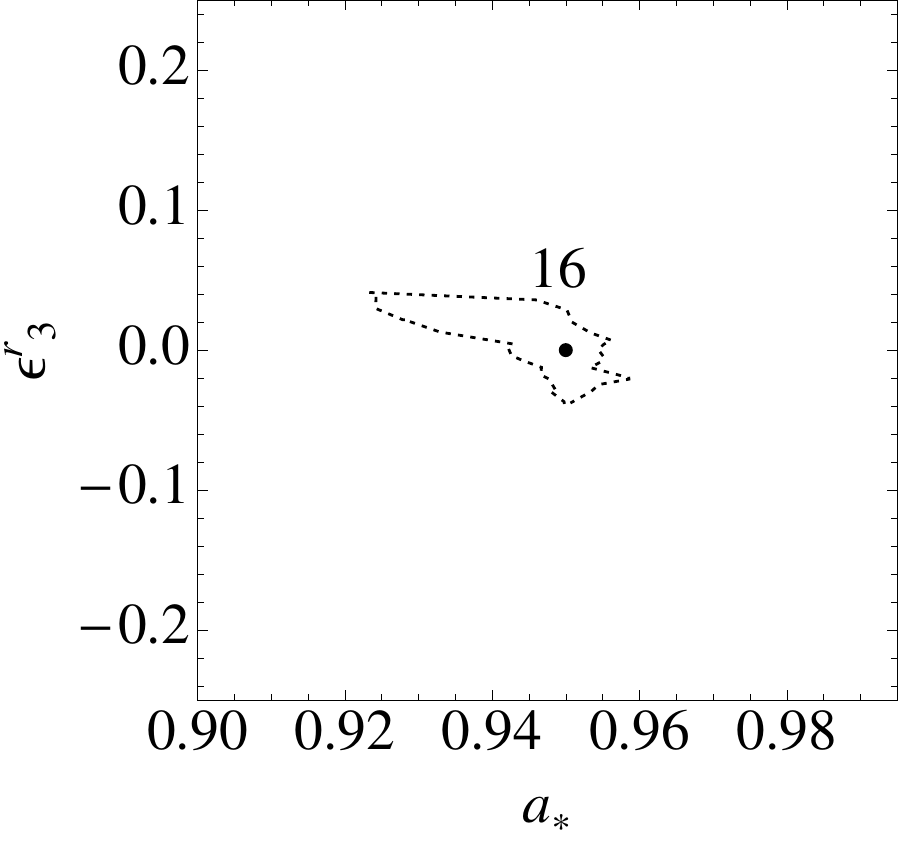}\\ 
\vspace{0.8cm}
\includegraphics[type=pdf,ext=.pdf,read=.pdf,width=7.0cm]{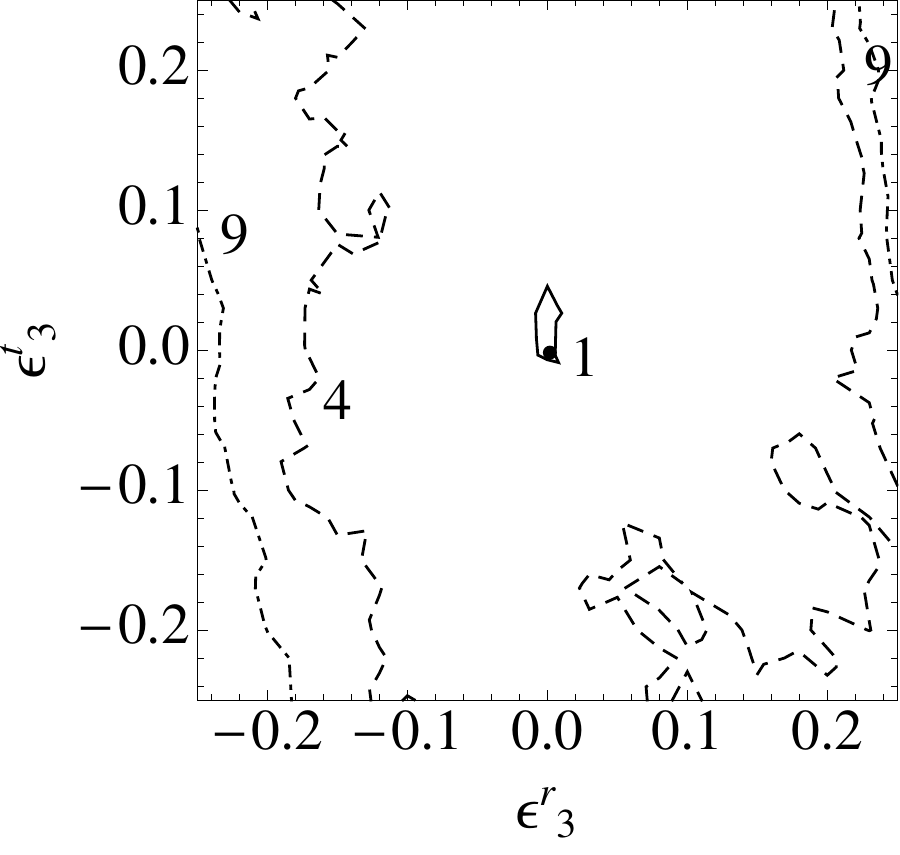}
\hspace{0.8cm}
\includegraphics[type=pdf,ext=.pdf,read=.pdf,width=7.0cm]{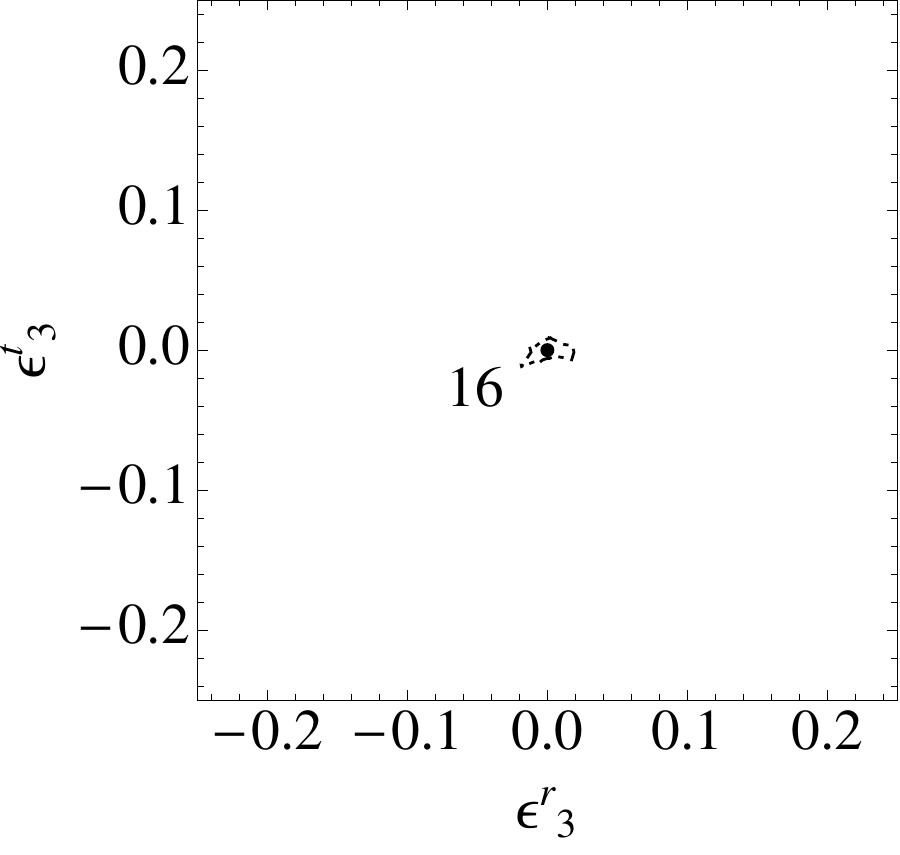} 
\end{center}
\caption{$\Delta\chi^2$ contours with $N_{\rm line} = 10^4$ photons from the analysis of the iron line profile (left panels) and the 2D transfer function (right panels). The reference model is a Kerr BH with spin parameter $a_*' = 0.95$ and inclination angle $i' = 70^\circ$. The top panels show the plots of $\Delta\chi^2$ of $a_*$ vs $\epsilon^t_3$ marginalized over $\epsilon^r_3$. The central panels show the plots of $\Delta\chi^2$ of $a_*$ vs $\epsilon^r_3$ marginalized over $\epsilon^t_3$. The bottom panels show the plots of $\Delta\chi^2$ of $\epsilon^t_3$ vs $\epsilon^r_3$ marginalized over $a_*$. The height of the source $h$, the ratio between the continuum and the iron line photon flux, $K$, and the photon index of the continuum, $\Gamma$, are also left as fit parameters. See the text for more details.}
\label{fig5}
\end{figure*}

The results of our simulations are shown in Figs.~\ref{fig1}-\ref{fig5}, where the time-integrated constraints are obtained from the same set of simulation data (using time-tagging) of the reverberation measurements. In Figs.~\ref{fig1} and \ref{fig2}, the reference model is a Kerr BH with $a_*' = 0.95$ and $i' = 70^\circ$. The high spin parameter and the high inclination angle make this case an ideal source to test the Kerr metric, because both serve to maximize the relativistic effects. Fig.~\ref{fig1} shows the time-integrated iron line measurement, and Fig.~\ref{fig2} gives the corresponding reverberation measurement. In both, the top panels correspond to a total photon count in the iron line $N_{\rm line} =10^3$, and the bottom panels to the case $N_{\rm line} =10^4$. In the left panels, we assume $\epsilon^r_3 = 0$ and fit for $\epsilon^t_3$. In the right panels, we consider the converse, fixing $\epsilon^t_3 = 0$ and treating $\epsilon^r_3$ as a fit parameter. It is evident that: $i)$ the reverberation measurement always produces stronger constraints than the corresponding time-integrated observation with the same $N_{\rm line}$, and $ii)$ while the time-integrated measurement cannot constrain $\epsilon^r_3$ even in the case $N_{\rm line} = 10^4$ (BHs with very large values of $\epsilon^r_3$ are allowed), the reverberation measurement is strongly constraining. Actually, the reverberation constraints for $N_{\rm line} = 10^4$ are so strong that systematic effects would certainly dominate given present data and limited model capabilities, which would impede any current measurement efforts.

Figs.~\ref{fig3} and \ref{fig4} show the analysis of a less favorable case, in the sense that the reference model is a Kerr BH with spin parameter $a_*' = 0.5$ observed from an inclination angle $i' = 20^\circ$. The low values of both the spin parameter and the inclination angle reduce the impact of the relativistic effect on the spectra of this source and therefore the constraints are weaker. Fig.~\ref{fig3} shows the contour levels of $\Delta\chi^2$ in the case of a time-integrated iron line measurement, Fig.~\ref{fig4} displays the corresponding reverberation constraints. As in Figs.~\ref{fig1} and \ref{fig2}, top panels refer to the case with a photon number count $N_{\rm line} = 10^3$, the bottom panels to the case with $N_{\rm line} =10^4$. In the left panels we consider the deformation parameter $\epsilon^t_3$ and $\epsilon^r_3 =0$, in the right panels $\epsilon^t_3=0$ and $\epsilon^r_3$ is free. The most important upshot is that the reverberation measurement with $N_{\rm line} =10^4$ seems to constrain $\epsilon^r_3$ even given low values of the spin parameter and of the inclination angle.

Lastly, Fig.~\ref{fig5} shows the case in which both $\epsilon^t_3$ and $\epsilon^r_3$ are allowed to be non-vanishing. The free parameters are thus seven ($a_*$, $\epsilon^t_3$, $\epsilon^r_3$, $i$, $h$, $\Gamma$, $K$). The left panels show the constraints from the time-integrated data, while the right panels show the case of reverberation measurements. The reference model is a Kerr black hole with spin parameter $a_*' = 0.95$ and observed from an inclination angle $i' = 70^\circ$. The photon count in the iron line is $N_{\rm line} = 10^4$. As in the previous plots, we have used the same set of simulation data to obtain the time-integrated and time-resolved constraints. The superior constraining power of a reverberation measurement is clear.

\section{Summary and conclusions \label{s-4}}

Astrophysical BH candidates are an ideal laboratory to test general relativity in the strong field regime. According to Einstein's theory of gravity, the spacetime geometry around these objects should be well described by the Kerr solution. The electromagnetic radiation emitted by  gas in the inner part of the accretion disk is affected by relativistic effects and the study of narrow (line) features in the spectrum can thus provide useful information for probing the background metric and testing the Kerr nature of BH candidates.

Parameter degeneracy typically hinders tests of the Kerr metric of the spacetime around BH candidates, in the sense that the same spectral features of a Kerr BH can be reproduced by a non-Kerr object with a different spin parameter. As shown by previous studies, the (time-integrated) iron K$\alpha$ line has the capability of breaking parameter degeneracy.  At the same time, some deviations from the Kerr geometry are more difficult to constrain via the line's profile than others. In Ref.~\cite{jjc2}, we found that the CPR deformation parameter $\epsilon^t_3$ is relatively easy to constrain, while it is much more challenging to constrain $\epsilon^r_3$ and even the iron line profile of a fast-rotating Kerr BH observed from a large inclination angle can be reproduced by a slow-rotating non-Kerr object with a large positive $\epsilon^r_3$.

In the present paper, we have investigated whether iron line reverberation measurements can better constrain the Kerr metric than the standard time-integrated iron line, in particular the CPR deformation parameter $\epsilon^r_3$. Such possibility was motivated by the fact that $\epsilon^r_3$ affects the photon propagation and therefore time information is naturally advantageous.  This is indeed what we find.

The results of our simulations are shown in Figs.~\ref{fig1}-\ref{fig5}. In the case of $N_{\rm line} =10^3$ photon counts in the iron line, corresponding to a typical long observation of a bright AGN with current X-ray facilities, the reverberation measurement can already provide a somewhat stronger constraint than the time-integrated observation, but it is still difficult to determine the CPR deformation parameter $\epsilon^r_3$. This is true even in the favorable situation of a fast-rotating Kerr BH observed from a high inclination angle. Its 2D transfer function cannot be distinguished from that of a non-Kerr object with different spin and large positive $\epsilon^r_3$.

In the case of an order-of-magnitude gain in photon count, $N_{\rm line} =10^4$ in our simulations, the time information of reverberation mapping provides a great advantage over the standard time-integrated analysis. Now we can constrain the CPR deformation parameter $\epsilon^r_3$, and this is true even without an ideal source. Even the 2D transfer function of a slow-rotating Kerr BH observed from a low inclination angle can be distinguished from that produced in the spacetime of deformed objects, see the bottom right panel in Fig.~\ref{fig4}. Iron line reverberation mapping is a more powerful technique for testing the Kerr metric and its merit with respect to a time-integrated approach improves drastically with photon count.  Such behavior is understood and results from the diminished impact of shot noise with signal strength when subdividing the spectrum in time. With sufficient line flux, it is possible to extract more information from the reverberation measurement on the spacetime geometry.


\begin{acknowledgments}
J.J. and C.B. were supported by the NSFC grants No.~11305038 and No.~U1531117, the Shanghai Municipal Education Commission grant No.~14ZZ001, the Thousand Young Talents Program, and Fudan University. C.B. also acknowledges support from the Alexander von Humboldt Foundation.
J.F.S. was supported by the NASA Einstein Fellowship grant PF5-160144.
\end{acknowledgments}


\end{document}